\documentstyle[aps,prl,epsfig,twocolumn,floats]{revtex}

\def\beq#1{\begin{equation}\label{#1}}
\def\eeq{\end{equation}}
\def\beqa#1{\begin{eqnarray}\label{#1}}
\def\eeqa{\end{eqnarray}}

\begin{document}

\draft

\tighten

\twocolumn[\hsize\textwidth\columnwidth\hsize\csname@twocolumnfalse\endcsname

\title{The Small-Scale Power Spectrum of Cold Dark Matter}

\author{Abraham Loeb$^\star$ \& Matias Zaldarriaga$^{\star,\dagger}$} 

\address{$\star$ Astronomy Department, Harvard University, 60 Garden
Street, Cambridge, MA 02138\\ $\dagger$ Physics
Department, Harvard University, 17 Oxford Street, Cambridge MA 02138}

\date{\today} \maketitle

\begin{abstract}

One of the best motivated hypotheses in cosmology states that most of the
matter in the universe is in the form of weakly-interacting massive
particles that decoupled early in the history of the universe and cooled
adiabatically to an extremely low temperature.  Nevertheless, the finite
temperature and horizon scales at which these particles decoupled imprint
generic signatures on their small scale density fluctuations. We show that
the previously recognized cut-off in the fluctuation power-spectrum due to
free-streaming of particles at the thermal speed of decoupling, is
supplemented by acoustic oscillations owing to the initial coupling between
the cold dark matter (CDM) and the radiation field. The power-spectrum
oscillations appear on the scale of the horizon at kinematic decoupling
which corresponds to a mass scale of $\sim 10^{-4} (T_{\rm d}/10~{\rm
MeV})^{-3}~M_\odot$ for a CDM decoupling temperature $T_{\rm d}$.  The
suppression of the power-spectrum on smaller scales by the acoustic
oscillations is physically independent from the free-streaming effect,
although the two cut-off scales are coincidentally comparable for $T_{\rm
d}\sim 10~{\rm MeV}$\ and a particle mass of $M\sim 100~{\rm GeV}$. The
initial conditions for recent numerical simulations of the earliest and
smallest objects to have formed in the universe, need to be modified
accordingly.  The smallest dark matter clumps may be detectable through
$\gamma$-ray production from particle annihilation, through fluctuations in
the event rate of direct detection experiments, or through their tidal
gravitational effect on wide orbits of objects near the outer edge of the
solar system.

\end{abstract}

]

\narrowtext

\paragraph*{Introduction.}

A broad range of observational data involving the dynamics of galaxies, the
growth of large-scale structure, and the dynamics and nucleosynthesis of
the universe as a whole, indicate the existence of dark matter with a mean
cosmic mass density that is $\sim 5$ times larger than the density of the
baryonic matter \cite{Jungman,WMAP}.  The data is consistent with a dark
matter composed of weakly-interacting, massive particles, that decoupled
early and adiabatically cooled to an extremely low temperature by the
present time \cite{Jungman}.  The Cold Dark Matter (CDM) has not been
observed directly as of yet, although laboratory searches for particles
from the dark halo of our own Milky-Way galaxy, have been able to restrict
the allowed parameter space for these particles. Since an alternative
more-radical interpretation of the dark matter phenomenology involves a
modification of gravity \cite{Beken}, it is of prime importance to find
direct fingerprints of the CDM particles. One such fingerprint involves the
small-scale structure in the universe \cite{Green}, on which we focus in
this paper.

Perhaps the most popular candidate for the CDM particle is a Weakly
Interacting Massive Particle (WIMP).  The lightest supersymmetric
particle (LSP) could be a WIMP (for a review see \cite{Jungman}).  The CDM
particle mass depends on free parameters in the particle physics model but
typical values cover a range around $M\sim 100~{\rm GeV}$ (up to values
close to a TeV). In many cases the LSP hypothesis will be tested at the
Large Hadron Collider (e.g. \cite{battaglia}) or in direct detection
experiments (e.g. \cite{baltz}).

The properties of the CDM particles affect their response to the
small-scale primordial inhomogeneities produced during cosmic
inflation. The particle cross-section for scattering off standard model
fermions sets the epoch of their thermal and kinematic decoupling from the
cosmic plasma (which is significantly later than the time when their
abundance freezes-out at a temperature $T\sim M$). Thermal decoupling is
defined as the time when the temperature of the CDM stops following that of
the cosmic plasma while kinematic decoupling is defined as the time when
the bulk motion of the two species start to differ.  For CDM the epochs of
thermal and kinetic decoupling coincide.  They occur when the time it takes
for collisions to change the momentum of the CDM particles equals the
Hubble time.  The particle mass determines the thermal spread in the speeds
of CDM particles, which tends to smooth-out fluctuations on very small
scales due to the free-streaming of particles after kinematic decoupling
\cite{Green,Green2}.  Viscosity has a similar effect before the CDM fluid
decouples from the cosmic radiation fluid \cite{Hofmann}.  An important
effect that has been previously ignored involves the memory the CDM fluid
has of the acoustic oscillations of the cosmic radiation fluid out of which
it decoupled.  Here we consider the imprint of these acoustic oscillations
on the small-scale power spectrum of density fluctuations in the universe.
Other imprints of acoustic oscillations on much larger scales were
identified recently in maps of the Cosmic Microwave Background (CMB)
\cite{WMAP}, and the distribution of nearby galaxies \cite{Eisenste}.
 
Throughout this paper, we adopt the standard set of cosmological parameters
\cite{WMAP} for a universe dominated by cold dark matter and a cosmological
constant ($\Lambda$--CDM).

\paragraph*{Formalism}

Kinematic decoupling of CDM occurs during the radiation-dominated era.  For
example, if the CDM is made of neutralinos with a particle mass of $\sim
100~{\rm GeV}$, then kinematic decoupling occurs at a cosmic temperature of
$T_{\rm d}\sim 10~{\rm MeV}$ \cite{Hofmann,Chen}.  As long as $T_d \ll
100~{\rm MeV}$, we may ignore the imprint of the QCD phase transition on
the CDM power spectrum \cite{Schmid}.  Over a short period of time during
this transition, the sound speed of the plasma vanishes, resulting in a
significant growth for perturbations with periods shorter than the length
of time over which the sound speed vanishes. The transition occurs when the
temperature of the cosmic plasma is $\sim 100-200~{\rm MeV}$ and lasts for
a small fraction of the Hubble time. As a result, the induced modifications
are on scales smaller than those we are considering here and the imprint of
the QCD phase transition is washed-out by the effects we calculate.

At early times the contribution of the dark matter to the energy density is
negligible. Only at relatively late times when the cosmic temperature drops
to values as low as $\sim 1$ eV, matter and radiation have comparable
energy densities.  As a result, the dynamics of the plasma at earlier times
is virtually unaffected by the presence of the dark matter particles.  In
this limit, the dynamics of the radiation determines the gravitational
potential and the dark matter just responds to that potential.  We will use
this simplification to obtain analytic estimates for the behavior of the
dark matter transfer function.

The primordial inflationary fluctuations lead to acoustic modes in the
radiation fluid during this era. The interaction rate of the particles in
the plasma is so high that we can consider the plasma as a perfect fluid
down to a comoving scale,
\begin{equation}
\lambda_f \sim \eta_{d}/\sqrt{N} \ \ \ \ ;  \ \ \ N \sim n \sigma t_d  ,
\end{equation}
where $\eta_d=\int_{0}^{t_d} dt/a(t)$ is the conformal time (i.e.  the
comoving size of the horizon) at the time of CDM decoupling, $t_d$;
$\sigma$ is the scattering cross section and $n$ is the relevant particle
density.  (Throughout the paper we set the speed of light and Planck's
constant to unity.) The damping scale depends on the species being
considered and its contribution to the energy density, and is the
largest for neutrinos which are only coupled through weak interactions. In
that case $N\sim (T/T_{d}^\nu)^3$ where $T_{d}^\nu\sim 1\ {\rm MeV}$ is the
temperature of neutrino decoupling. At the time of CDM decoupling $N\sim
M/T_d \sim 10^4$ for the rest of the plasma, where $M$ is the mass of the
CDM particle.  In this paper we will consider modes of wavelength larger
than $\lambda_f$, and so we neglect the effect of radiation diffusion
damping and treat the plasma (without the CDM) as a perfect fluid.

The equations of motion for a perfect fluid during the radiation era can be
solved analytically. We will use that solution here, following the notation
of Ref. \cite{Dodelson}.  As usual we Fourier decompose fluctuations and
study the behavior of each Fourier component separately. For a mode of
comoving wavenumber $k$ in Newtonian gauge, the gravitational potential
fluctuations are given by:
\begin{equation}
\Phi= 3\Phi_p\left[{\sin(\omega\eta) -\omega\eta\cos(\omega\eta)
\over (\omega\eta)^3}\right],
\label{eq:phi}
\end{equation}
where $\omega=k/\sqrt{3}$ is the frequency of a mode and $\Phi_p$ is its
primordial amplitude in the limit $\eta \rightarrow 0$. In this paper we
use conformal time $\eta=\int dt/a(t)$ with $a(t)\propto t^{1/2}$ during
the radiation-dominated era. The monopole $\Theta_0$ and dipole $\Theta_1$
of the photon distribution can be written in terms of the gravitational
potential as:
\begin{eqnarray}
\Theta_0&=&\Phi\left({x^2\over 6}+{1\over 2}\right)+{x\over 2}\Phi^\prime
\nonumber \\ \Theta_1&=&-{x^2\over 6}\left(\Phi^\prime +{1\over x}
\Phi\right)
\label{eq:theta01}
\end{eqnarray}
where $x\equiv k\eta$ and a prime denotes a derivative with respect to $x$.

The solutions in equations (\ref{eq:phi}) and (\ref{eq:theta01}) assume
that both the sound speed and the number of relativistic degrees of freedom
are constant over time. As a result of the QCD phase transition and of
various particles becoming non-relativistic, both of these assumptions are
not strictly correct. The vanishing sound speed during the QCD phase
transition provides the most dramatic effect, but its imprint is on scales
smaller than the ones we consider here because the transition occurs at a
significantly higher temperature and only lasts for a fraction of the
Hubble time \cite{Schmid}.

Before the dark matter decouples kinematically we will treat it as a fluid
which can exchange momentum with the plasma through particle collisions. At
early times, the CDM fluid follows the motion of the plasma and is involved
in its acoustic oscillations. The continuity and momentum equations for the
CDM can be written as:
\begin{eqnarray}
\dot \delta_c+\theta_c &=& 3 \dot \Phi \nonumber \\ \dot \theta_c + {\dot a
\over a} \theta_c &=& k^2 c_s^2 \delta_c - k^2 \sigma_c - k^2 \Phi +
\tau_c^{-1} (\Theta_1 - \theta_c)
\label{eq:dmcontmom}
\end{eqnarray}
where a dot denotes an $\eta$-derivative, $\delta_c$ is the dark matter density
perturbation, $\theta_c$ is the divergence of the dark matter velocity
field and $\sigma_c$ denotes the anisotropic stress. In writing these
equations we have followed Ref. \cite{Ma}. The term $\tau_c^{-1} (\Theta_1 -
\theta_c)$ encodes the transfer of momentun between the radiation and CDM
fluids and $\tau_c^{-1}$ provides the collisional rate of momentum transfer,
\begin{equation}
\tau_c^{-1}= n \sigma {T \over M} a, 
\end{equation}  
with $n$ being the number density of particles with which the dark matter
is interacting, $\sigma(T)$ the average cross section for interaction and
$M$ the mass of the dark matter particle. The relevant scattering partners
are the standard model leptons which have thermal abundances.  For detailed
expressions of the cross section in the case of supersymmetric (SUSY) dark
matter, see Refs. \cite{Chen,Green2}. For our purpose, it is sufficient to
specify the typical size of the cross section and its scaling with cosmic
time,
\begin{equation}
\sigma\approx {T^2 \over M_\sigma^4}  ,
\end{equation}
where the coupling mass $M_\sigma$ is of the order of the weak-interaction
scale ($\sim 100$ GeV) for SUSY dark matter. This equation should be taken
as the definition of $M_\sigma$, as it encodes all the uncertainties in the
details of the particle physics model into a single parameter. The
temperature dependance of the averaged cross section is a result of the
available phase space. Our results are quite insensitive to the details
other than through the decoupling time. Equating $\tau_c^{-1}/a$ to the
Hubble expansion rate gives the temperature of kinematic decoupling:
\begin{equation}
T_d= \left({M_\sigma^4 M \over M_{pl}}\right)^{1/4}\approx 10 \ {\rm MeV}
\left({M_\sigma \over 100 \ {\rm GeV}}\right) \left( M \over 100\ {\rm GeV}
\right)^{1/4}  .
\end{equation}

The term $k^2 c_s^2 \delta_c$ in Eq. (\ref{eq:dmcontmom}) results from the
pressure gradient force and $c_s$ is the dark matter sound speed.  In the
tight coupling limit, $\tau_c \ll H^{-1}$ we find that $c_s^2\approx f_c
T/M$ and that the shear term is $k^2 \sigma_c \approx f_v c_s^2 \tau_c
\theta_c$.  Here $f_{v}$ and $f_{c}$ are constant factors of order unity.
We will find that both these terms make a small difference on the scales of
interest, so their precise value is unimportant.

By combining both equations in (\ref{eq:dmcontmom}) into a single equation
for $\delta_c$ we get
\begin{eqnarray}
\delta_c^{\prime\prime}&+& {1\over x}\left[1+ F_{\rm
v}(x)\right]\delta_c^\prime +c_s^2(x) \delta_c \nonumber \\ = &S(x)&
-3F_{\rm v}(x)\Phi^\prime+{x_d^4 \over x^5} \left(3\theta_0^\prime
-\delta_c^\prime\right),
\label{eq:delta}
\end{eqnarray}
where $x_d=k\eta_d$ and $\eta_d$ denotes the time of kinematic decoupling
which can be expressed in terms of the decoupling temperature as,
\begin{eqnarray}
\eta_d= {2 t_d (1+z_d) }\approx {M_{pl} \over T_0 T_d} &\approx& 10 \ {\rm
pc} \ \left({T_d \over 10\ {\rm MeV}}\right)^{-1} \nonumber \\ &\propto&
M_\sigma^{-1} M^{-1/4} ,
\end{eqnarray}
with $T_0=2.7$K being the present-day CMB temperature and $z_d$ being the
redshift at kinematic decoupling.  We have also introduced the source
function,
\begin{equation}
S(x)\equiv -3\Phi^{\prime\prime}+\Phi-{3\over x}\Phi^\prime.
\label{eq:s}
\end{equation}
For $x\ll x_d$, the dark matter sound speed is given by 
\begin{equation}
c_s^2(x)=c_s^2(x_d) {x_d\over x}, 
\label{eq:sound}
\end{equation}
where $c_s^2(x_d)$ is the dark matter sound speed at kinematic decoupling
(in units of the speed of light),
\begin{equation}
c_s(x_d) \approx 10^{-2} f_c^{1/2} \left({T_d \over 10\ {\rm MeV}}
\right)^{1/2} \left({M \over 100\ {\rm GeV}} \right)^{-1/2}.
\end{equation}
In writing (\ref{eq:sound}) we have assumed that prior to decoupling the
temperature of the dark matter follows that of the plasma. For the
viscosity term we have,
\begin{equation}
F_{v}(x)= f_{v} c_s^2(x_d) x_d^2  \left({x_d\over x}\right)^5.
\label{eq:F_v_before}
\end{equation}

\paragraph*{Free streaming after kinematic decoupling}

In the limit of the collision rate being much slower than the Hubble
expansion, the CDM is decoupled and the evolution of its perturbations is
obtained by solving a Boltzman equation: \begin{equation} {\partial f \over
\partial \eta} +{d x_i \over d\eta} {\partial f \over \partial x_i} + {d
q_i \over d\eta} {\partial f \over \partial q_i} =0,
\end{equation}
where $f(\vec x,\vec q,\eta)$ is the distribution function which depends on
time, position and comoving momentum $\vec q$. The comoving momentum
3-components are ${d x_i / d\eta} = q_i/a$. We use the Boltzman equation to
find the evolution of modes that are well inside the horizon with $x\gg
1$. In the radiation era, the gravitational potential decays after horizon
crossing (see Eq. \ref{eq:phi}). In this limit the comoving momentum
remains constant, ${d q_i /d\eta} =0$ and the Boltzman equation becomes,
\begin{equation}
{\partial f \over \partial \eta} +{q_i \over a}  {\partial f \over \partial x_i}  =0. 
\end{equation}
We consider a single Fourier mode and write $f$ as, 
\begin{equation}
f(\vec x,\vec q,\eta)=f_0(q) [1+\delta_F(\vec q,\eta) e^{i \vec k \cdot
\vec x}],
\end{equation}
where $f_0(q)$ is the unperturbed distribution,
\begin{equation}
f_0(q) = n_{\rm CDM} \left( {M \over 2\pi T_{\rm CDM} } \right)^{3/2}
\exp\left[-{1\over 2} {M q^2 \over T_{\rm CDM}}\right]
\end{equation} 
where $n_{\rm CDM}$ and $T_{\rm CDM}$ are the present-day density and
temperature of the dark matter.

Our approach is to solve the Boltzman equation with initial conditions
given by the fluid solution at a time $\eta_*$ (which will depend on
$k$). The simplified Boltzman equation can be easily solved to give
$\delta_F(\vec q,\eta)$ as a function of the initial conditions
$\delta_F(\vec q,\eta_*)$,
\begin{equation}
\delta_F(\vec q,\eta)=\delta_F(\vec q,\eta_*) \exp[-i \vec q \cdot \vec k
{\eta_* \over a(\eta_*)} \ln (\eta/\eta_*)] .
\label{solbol}
\end{equation}   

The CDM overdensity $\delta_c$ can then be expressed in terms of the
perturbation in the distribution function as,
\begin{equation}
\delta_c(\eta)={1\over n_{\rm CDM}} \int d^3q \ f_0(q) \ \delta_F(\vec q,\eta).
\end{equation} 
We can use (\ref{solbol}) to obtain the evolution of $\delta_c$ in terms of
its value at $\eta_*$,
\begin{equation}
\delta_c(\eta)= \exp\left[-{1\over 2} {k^2 \over k_f^2} \ln^2({\eta\over
\eta_*})\right] \ \left[\delta|_{\eta_*} + {d \delta \over d
\eta}|_{\eta_*} \eta_* \ln({\eta\over \eta_*})\right] ,
\label{delcbol}
\end{equation}
where $k_f^{-2}= \sqrt{(T_d/ M)} \eta_d$. The exponential term is
responsible for the damping of perturbations as a result of free streaming and
the dispersion of the CDM particles after they decouple from the plasma. The
above expression is only valid during the radiation era. The free streaming
scale is simply given by $\int dt (v/a) \propto \int dt a^{-2}$ which grows
logarithmically during the radiation era as in equation (\ref{delcbol}) but
stops growing in the matter era when $a\propto t^{2/3}$.

Equation (\ref{delcbol}) can be used to show that even during the free
streaming epoch, $\delta_c$ satisfies equation (\ref{eq:delta}) but with a
modified sound speed and viscous term. For $x \gg x_d$ one should use,
\begin{eqnarray}
c_s^2(x) &=& c_s^2(x_d) \left({x_d\over x}\right)^2 \left[1+ x_d^2
c_s^2(x_d) \ln^2({x\over x_d})\right] \nonumber \\ F_{v}(x)&=&2 c_s^2(x_d)
x_d^2 \ln\left({x_d\over x}\right)
\end{eqnarray}
The differences between the above scalings and those during the tight
coupling regime are a result of the fact that the dark matter temperature
stops following the plasma temperature but rather scales as $a^{-2}$ after
thermal decoupling, which coincides with the kinematic decoupling.
We ignore the effects of heat transfer during the
fluid stage of the CDM because its temperature is controlled by the much
larger heat reservoir of the radiation-dominated plasma at that stage.

To obtain the transfer function we solve the dark matter fluid equation
until decoupling and then evolve the overdensity using equation
(\ref{delcbol}) up to the time of matter--radiation equality. In practice,
we use the fluid equations up to $x_*=10\  {\rm max}(x_d, 10)$ so as to switch
into the free streaming solution well after the gravitational potential has
decayed. In the fluid equations, we smoothly match the sound speed and
viscosity terms at $x = x_d$. As mentioned earlier, because $c_s(x_d)$ is
so small and we are interested in modes that are comparable to the size of
the horizon at decoupling, i.e. $x_d \sim {\rm few}$, both the dark
matter sound speed and the associated viscosity play only a minor
role, and our simplified treatment is adequate.

\begin{figure}[th]
\centerline{\epsfig{file=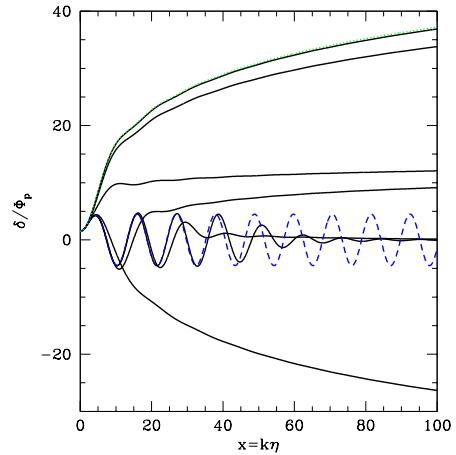,height=2.5in}}
\caption{The normalized amplitude of CDM fluctuations $\delta/\Phi_P$ for a
variety of modes with comoving wavenumbers
$\log(k\eta_d)=(0,1/3,2/3,1,4/3,5/3,2)$ as a function of $x\equiv k\eta$,
where $\eta=\int_0^t dt/a(t)$ is the conformal time coordinate. The dashed
line shows the temperature monopole $3\theta_0$ and the dotted curve shows
the evolution of a mode that is uncoupled to the cosmic plasma.}
\label{figtime}
\end{figure}

In Figure \ref{figtime} we illustrate the time evolution of modes during
decoupling for a variety of $k$ values. The situation is clear. Modes that
enter the horizon before kinematic decoupling oscillate with the
radiation fluid. This behavior has two important effects. In the absence of
the coupling, modes receive a ``kick" by the source term $S(x)$ as they
cross the horizon. After that they grow logarithmically. In our case, modes
that entered the horizon before kinematic decoupling follow the plasma
oscillations and thus miss out on both the horizon ``kick" and the
beginning of the logarithmic growth. Second, the decoupling from the
radiation fluid is not instantaneous and this acts to further damp the
amplitude of modes with $x_d \gg 1$. This effect can be understood as
follows.  Once the oscillation frequency of the mode becomes high compared
to the scattering rate, the coupling to the plasma effectively damps the
mode. In that limit one can replace the forcing term $\Theta_0^\prime$ by
its average value, which is close to zero. Thus in this regime, the
scattering is forcing the amplitude of the dark matter oscillations to
zero.  After kinematic decoupling the modes again grow logarithmically but
from a very reduced amplitude. {\it The coupling with the plasma induces
both oscillations and damping of modes that entered the horizon before
kinematic decoupling. This damping is different from the free streaming
damping that occurs after kinematic decoupling}.

\begin{figure}[th]
\centerline{\epsfig{file=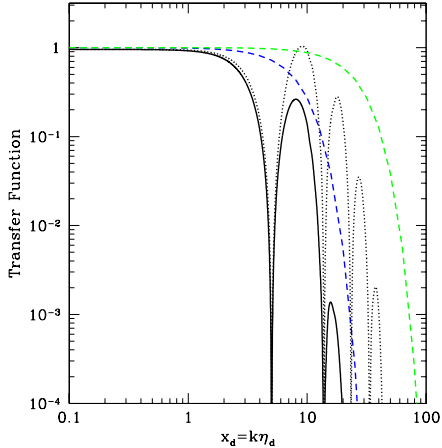,height=2.5in}}
\caption{Transfer function of the CDM density perturbation amplitude
(normalized by the primordial amplitude from inflation).  We show two
cases: {\it (i)} $T_d/M=10^{-4}$ and $T_d/T_{\rm eq}=10^7$; {\it (ii)}
$T_d/M=10^{-5}$ and $T_d/T_{\rm eq}=10^7$. In each case the oscillatory
curve is our result and the other curve is the free-streaming only result
that was derived previously in the literature [4,7,8].  }
\label{figtransfer}
\end{figure}

In Figure \ref{figtransfer} we show the resulting transfer function of the
CDM overdensity. The transfer function is defined as the ratio between the
CDM density perturbation amplitude $\delta_c$ when the effect of the
coupling to the plasma is included and the same quantity in a model where
the CDM is a perfect fluid down to arbitrarily small scales (thus, the
power spectrum is obtained by multiplying the standard result by the square
of the transfer function).  This function shows both the oscillations and
the damping signature mentioned above. The peaks occur at multipoles of the
horizon scale at decoupling,
\begin{equation} k_{peak}= (8,15.7,24.7,..) \eta_d^{-1} \propto {M_{pl} \over
T_0 T_d}.
\end{equation} This same scale determines the ``oscillation" damping. The
free streaming damping scale is,
\begin{equation}
\eta_d c_d(\eta_d) \ln(\eta_{eq}/\eta_d) \propto {M_{pl} M^{1/2} \over T_0
T_d^{3/2}} \ln(T_d/T_{\rm eq}),
\end{equation}
where $T_{\rm eq}$ is the temperature at matter radiation equality,
$T_{\rm eq}\approx 1\ {\rm eV}$. The free streaming scale is parametrically
different from the ``oscillation" damping scale. However for our fiducial
choice of parameters for the CDM particle they roughly coincide.

The vanishing of the sound speed during the QCD phase transition amplifies
perturbations which have $\omega \Delta \eta_{QCD}>1$, where $\Delta
\eta_{QCD}$ is the duration of the transition. In Figure \ref{figtransfer}
the affected modes are those with $x_d=k \eta_d > \sqrt{3} \ (\eta_{QCD}/
\Delta \eta_{QCD}) \ (\eta_d/\eta_{QCD})$. Typical values of
$\eta_d/\eta_{QCD}\sim 10-15$ and $\eta_{QCD}/ \Delta \eta_{QCD}\sim 3-10$
relate this condition to modes with $x_d > 50-260$. Thus the affected
scales are severely damped by the effects considered in this paper.

Finally we want to stress the fact that the damping scale is significantly
smaller than the scales observed directly in the Cosmic Microwave
Background or through large scale structure surveys. For example, the ratio
of the damping scale to the scale that entered the horizon at equality is
$\eta_{d}/\eta_{eq}\sim T_{eq}/T_d \sim 10^{-7}$ and to our present horizon
$\eta_{d}/\eta_{0}\sim (T_{eq} T_0)^{1/2}/T_{d} \sim 10^{-9}$. In the
context of inflation, these scales were created 16 and 20 {\it e}--folds
apart. Given the large extrapolation, one could certainly imagine that a
change in the spectrum could alter the shape of the power spectrum around
the damping scale.  However, for smooth inflaton potentials with small
departures from scale invariance this is not likely to be the case. On
scales much smaller than the horizon at matter radiation equality, the
spectrum of perturbations density before the effects of the damping are
included is approximately,
\begin{eqnarray}
\Delta^2(k)&\propto& \exp\left[(n-1)\ln(k\eta_{eq})+ {1\over 2} \alpha^2
\ln(k\eta_{eq})^2 + \cdots\right]\nonumber \\ && \times \ln^2(k\eta_{eq}/8)
\end{eqnarray}
where the first term encodes the shape of the primordial spectrum and the
second the transfer function. Primordial departures from scale invariance
are encoded in the slope $n$ and its running $\alpha$. The effective slope
at scale $k$ is then,
\begin{equation}
{\partial \ln \Delta^2 \over \partial \ln k}= (n-1)+\alpha \ln(k\eta_{eq})
+ {2 \over \ln(k\eta_{eq}/8)}.
\end{equation}
For typical values of $(n-1) \sim 1/60$ and $\alpha\sim 1/60^2$ the slope
is still positive at $k\sim \eta_d^{-1}$, so the cut-off in the power will
come from the effects we calculate rather than from the shape of the
primordial spectrum. However given the large extrapolation in scale, one
should keep in mind the possibility of significant effects resulting from
the mechanisms that generates the density perturbations.

\paragraph*{Implications}
We have found that acoustic oscillations, a relic from the epoch when the
dark matter coupled to the cosmic radiation fluid, truncate the CDM power
spectrum on a comoving scale larger than effects considered before, such as
free-streaming and viscosity \cite{Green,Green2,Hofmann}. For SUSY dark
matter, the minimum mass of dark matter clumps that form in the universe is
therefore increased by more than an order of magnitude to a value of
\footnote{Our definition of the cut-off mass follows the convention of the
Jeans mass, which is defined as the mass enclosed within a sphere of radius
$\lambda_{\rm J}/2$ where $\lambda_{\rm J}\equiv 2\pi/k_{\rm J}$ is the
Jeans wavelength \cite{Haiman}.}
\begin{eqnarray}
M_{\rm cut}&=& {4\pi\over 3} \left({\pi \over k_{\rm cut}}\right)^{3}
\Omega_M \rho_{\rm crit} \nonumber \\ &\simeq& 10^{-4} \left({T_d\over
10~{\rm MeV}}\right)^{-3} M_\odot,
\label{clump}
\end{eqnarray}
where $\rho_{\rm crit}=(H_0^2/8\pi G)=9\times 10^{-30}~{\rm g~cm^{-3}}$ is
the critical density today, and $\Omega_M$ is the matter density for the
concordance cosmological model \cite{WMAP}. We define the cut-off
wavenumber $k_{\rm cut}$ as the point where the transfer function first
drops to a fraction $1/e$ of its value at $k\rightarrow 0$. This
corresponds to $k_{\rm cut}\approx 3.3 \ \eta_d^{-1}$.

Recent numerical simulations \cite{Moore,Gao} of the earliest and smallest
objects to have formed in the universe \cite{BL01}, need to be redone for
the modified power spectrum that we calculated in this paper. Although it
is difficult to forecast the effects of the acoustic oscillations through
the standard Press-Schechter formalism \cite{Press}, it likely that the
results of such simulations will be qualitatively the same as before except
that the smallest clumps would have a mass larger than before (as given by
Eq. \ref{clump}).

Potentially, there are several observational signatures of the smallest CDM
clumps.  As pointed out in the literature \cite{Moore,Stoehr}, the smallest
CDM clumps could produce $\gamma$-rays through dark-matter annihilation in
their inner density cusps, with a flux in excess of that from nearby dwarf
galaxies.  If a substantial fraction of the Milky Way halo is composed of
CDM clumps with a mass $\sim 10^{-4} M_\odot$, the nearest clump is
expected to be at a distance of $\sim 4\times 10^{17}$ cm.  Given that the
characteristic speed of such clumps is a few hundred ${\rm km~s^{-1}}$, the
$\gamma$-ray flux would therefore show temporal variations on the
relatively long timescale of a thousand years.  Passage of clumps through
the solar system should also induce fluctuations in the detection rate of
CDM particles in direct search experiments.

Other observational effects have rather limited prospects for
detectability.  Because of their relatively low-mass and large size ($\sim
10^{17}~{\rm cm}$), the CDM clumps are too diffuse to produce any
gravitational lensing signatures (including {\it femto-}lensing
\cite{Gould}), even at cosmological distances. They could, however, have an
effect on the time delay of the signal from transient cosmological sources.
We illustrate this effect by considering a characteristic CDM clump with a
size $\sim 10^{17}$cm and a mass of ~$10^{-4}M_\odot$ (corresponding to
formation redshift $\sim 50$). If a major fraction of the intergalactic CDM
is in the form of these clumps (amounting to a mean mass density of $\sim
3\times 10^{-30}~{\rm g~cm^{-3}}$), then the mean separation between clumps
is $\sim 6\times 10^{19}$cm$=20$pc. The gravitational potential would then
fluctuate on this length scale everywhere in the universe with an amplitude
of $(\Phi/c^2)\sim 2\times 10^{-19}$.  Now, consider a compact source at
cosmological distances with a size comparable to the distance between
clumps, $\sim 10$pc.
Photons emerging from different regions of the source will encounter
different gravitational potential delays (the so-called {\it Shapiro time
delay} \cite{Shapiro}) along their path due to the fact that they traverse
CDM clumps at different impact parameters. For a source that is much larger
than $\sim 20$pc, the effect of multiple CDM clumps would average out
because the source would sample the full distribution of impact parameters
with small statistical fluctuations around the mean time delay.  The mean
{\it Shapiro delay} from the network of CDM clumps along the line-of-sight
to a cosmological source is $\sim (\Phi/c^2)t_H\sim 0.1$ sec, where $t_H$
is the light travel time across the observable universe. The total number
of clumps along the line-of-sight is $N\sim (4 {\rm Gpc/20pc})\sim 2\times
10^8$. This implies that the variance in time delays across a source of
size $\sim 10$pc would be suppressed by a factor of $\sim 1/\sqrt{N}\sim
10^{-4}$ relative to the mean time delay, i.e. it would have a magnitude
$\sim 10~{\rm \mu sec}$.  $\gamma$-ray bursts may possess variability on
such a timescale (which would have been smoothed out if their source had a
size $\sim 10$pc), but their estimated source size is $R<0.1$pc
\cite{Piran} and so the temporal smoothing effect for them is smaller by
another factor $>(20/0.1)=2\times 10^{2}$, which leads to a variance in the
delay timescale of less than a microsecond.  Unfortunately this time delay
appears unmeasurable, since the variability of any source of apparent
dimension $R$ would be expected to be smoothed-out on a timescale, $\sim
R/\gamma c$, where $\gamma$ is the Lorentz factor of the emitting material
(which is $<10^3$ for $\gamma$-ray bursts).

The smallest CDM clumps should not affect the intergalactic baryons which
have a much larger Jeans mass \cite{BL01}. However, once objects above
$\sim 10^6M_\odot$ start to collapse at redshifts $z<30$, the baryons would
be able to cool inside of them via molecular hydrogen transitions and the
interior baryonic Jeans mass would drop.  The existence of dark matter
clumps could then seed the formation of the first stars inside these
objects \cite{Bromm}.

Finally, we note that the smallest CDM clumps may have a dynamical effect
on wide orbits of test particles near the outer edge the solar system
after the effects of the known planets have been modeled to a high
precision \cite{Stein}.

\paragraph*{Acknowledgments.}

This work was supported in part by NASA grant NAG 5-13292, NSF grants
AST-0071019, AST-0204514 (for A.L.) and by NSF grants AST-0098606,
PHY-0116590 and the David \& Lucille Packard Foundation Fellowship (for
M.Z.).

\end{document}